\documentclass[conference]{IEEEtran}
\IEEEoverridecommandlockouts
\pdfoutput=1 
\usepackage[utf8]{inputenc}
\usepackage[T1]{fontenc}
\usepackage{amsmath,amssymb,amsfonts}
\usepackage{algorithmic}
\usepackage{graphicx}
\usepackage{textcomp}
\usepackage[table,xcdraw]{xcolor}
\usepackage{xcolor}
\usepackage{multirow}
\usepackage{tabularx,ragged2e,booktabs,caption}
\usepackage{biblatex} 
\usepackage{enumitem}
\usepackage{pifont}
\def\BibTeX{{\rm B\kern-.05em{\sc i\kern-.025em b}\kern-.08em
    T\kern-.1667em\lower.7ex\hbox{E}\kern-.125emX}}
\usepackage{xspace}
\usepackage{flushend}
\usepackage{algorithm2e}
\begin{document}
\RestyleAlgo{ruled}
\title{QuCS: A Lecture Series on Quantum Computer Software and System\\
\vspace{5pt}

\vspace{-12pt}
}

\newcommand{\HW}[1]{\textcolor{blue}{[HW: #1]}}

\author{\IEEEauthorblockN{
Zhiding Liang\textsuperscript{1}  \ \
Hanrui Wang\textsuperscript{2} }
\IEEEauthorblockA{
\textsuperscript{1}University of Notre Dame, IN, USA.
\textsuperscript{2}Massachusetts Institute of Technology,
MA, USA.
\vspace{-0.15in}}
}

\maketitle

\begin{abstract}
In this era of incessant advancements in quantum computing, bridging the gap between quantum algorithms' hardware requisites and available devices has become crucial. A prime focus in this context is the Software and System Level support for quantum computers, which has shown promising potential in significantly decreasing this gap. However, a noteworthy deficit of quantum software and system level-focused courses has been observed in academia worldwide. Addressing this deficiency, this paper proposes the Quantum Computer Systems (QuCS) Lecture Series. The QuCS Lecture Series aims to enhance the visibility of quantum computing software and system level and foster diverse participation in quantum computing research across multiple universities worldwide. It is envisioned as an inclusive platform to bring together individuals of diverse backgrounds, catalyzing cross-cultural collaboration and innovation in this burgeoning field.

The lecture series begins with an introductory session elucidating the core concepts and fundamentals of quantum computing. This foundational knowledge will be built upon in subsequent sessions, highlighting cutting-edge research trends and recent findings in quantum software and system level. This paper provides a comprehensive overview of the QuCS Lecture Series, detailing the format, the gamut of topics to be covered, and their significance. It emphasizes the potential impact of the series on accelerating progress towards quantum supremacy and fostering a diverse, global community of quantum computing researchers and practitioners. The QuCS Lecture Series has already hosted nearly 40 lectures with over 40 confirmed speakers from more than eight different countries and from both academia and industry, QuCS also attracted more than 1000 subscribers from all over the world.

\end{abstract}

\begin{IEEEkeywords}
Quantum Computing, Education, Quantum Workforce, Software, Computer Systems, Computer Architecture
\end{IEEEkeywords}

\section{Introduction}
\label{sec1}
Hardware developments for quantum computing have witnessed tremendous breakthroughs in the recent decade~\cite{zhan2022transmitter, hillery2023discrete, qi2023theoretical, maldonado2022quantum}, where different types of quantum bits have been developed, such as superconducting transmon, trapped-ion, and neutral atoms. 
Google released Sycamore~\cite{arute2019quantum} in 2019 and claimed to have achieved quantum supremacy by performing random circuit sampling on it. This problem is in principle difficult to simulate with classical computers, but it is far from being practically useful. 
In late 2021, IBM released a 127-qubit quantum computer with heavy-hexagon hardware topology~\cite{ibmq127qubits}. 
Quantum computers with about 1,000 qubits will be built in the next decade. 
Besides the breakthroughs in the aspect of quantum hardware, new quantum algorithms have also been proposed and demonstrated during the past years.
As a result, the field of quantum computing has attracted widespread attention from academia, industry, and governments. And the demand for quantum computing professionals has increased drastically.

The research topics of quantum computing were originally limited to mathematics and physics, but in the past decade were expanded to include chemistry, electrical engineering, and computer science. 
A growing number of companies have created online platforms that allow students and researchers to perform cloud quantum computing without having a real quantum computer locally, such as IBM's Qiskit~\cite{qiskit}, Google's Cirq~\cite{cirq}, AWS's Braket~\cite{amazon}, and D-Wave's Ocean~\cite{wave}.
Except for accessible real-machines, there are many open-source quantum simulators available such as Quandry~\cite{gunther2021quandary}, Qutip~\cite{johansson2012qutip}, Juqbox~\cite{petersson2020discrete}, and torchquantum~\cite{wang2022quantumnas}.

As we know, quantum noises form the bottleneck of current quantum machines and various research attempts have been made to mitigate the high noise of quantum devices in the Noisy Intermediate-Scale Quantum (NISQ) era. Some works focused on exploring error-resilient algorithms, while others seek solutions from error correction methods~\cite{liang2022variational, wang2021roqnn, wang2022quantumnas, wu2022synthesis, das2022afs, cheng2023fidelity}.
Recent works have also tried to reduce the compilation overhead of quantum programs~\cite{shi2020resource, cheng2020accqoc, baker2021virtual, jiang2021co,li2020towards,tannu2022hammer,das2022foresight, wang2022torchquantum, chen2022stable,liang2022hybrid, qin2022improving, cheng2022topgen}. All these efforts are made at the software and system level to improve the quantum computers.

In recent years, 
increasingly more universities started to offer courses on quantum computing.
Gatti and Sotelo offered quantum computing courses and reported the feedbacks from the Universidad de Montevideo~\cite{gatti2021quantum}. ~\cite{ibmonline, Uchicago, XPRO} offer online courses about quantum computing fundamentals with tuition fee.~\cite{salehi2022computer} presents a computer science-oriented approach to teach quantum computing. Angara et al.~\cite{angara2020quantum} provided quantum computing courses in high school, while Carrascal et al.~\cite{carrascal2021first} are targeting undergraduate students. Also,~\cite{mykhailova2020teaching} describes the experience of teaching a quantum computing course via practical software-driven approach. 
However, we notice that most of the courses only offer general introduction to quantum computing, neglecting the important role of software and system level support. Hence, we decided to hold lectures to introduce the software and system level support for quantum computing.

\subsection{Classification of Quantum Computing Fields}
In order to make clear the topics of our lecture series, we introduce here the classification of different fields in quantum computing. In general, the fields of quantum computing could be divided into three major categories:
developments of reliable quantum devices, improvements and theoretical analysis for quantum algorithms, and software and system support for quantum computers.
We can categorize the subdivision of research directions into these categories. Our lecture series mainly focus on the 
software and system support for quantum computing.

\subsection{Target Audience}
Initially, we found that the University of Notre Dame did not have a software and system level course on quantum computing, and some students asked us for open-source resources to learn. At that time. we often recommend the online courses offered by the University of Chicago ~\cite{Uchicago}, the quantum computing courses offered by MIT xPRO ~\cite{XPRO}, and the quantum computing summer school offered by IBM ~\cite{ibmonline}.
After investigations are conducted on the educational resources for quantum computing, we found that most universities do not offer courses on quantum computing at the software and system level. Therefore, we decided to organize a series of lectures, starting from basic introduction to cutting-edge research topics. The purpose is to provide more opportunities for the public to learn quantum computing.
Our target audience include but is not limited to, practitioners and researchers in quantum computing, students who are interested but new to the area, and quantum educators.

\subsection{About This Paper}

This paper presents an innovative lecture series launched in Spring 2022, with the inaugural lecture taking place on July 7th, 2022. Unlike traditional lecture series that offer a broad spectrum of topics, ours has a two-fold structure: an introductory session, followed by a cutting-edge research session.

The lecture series is unique as it concentrates solely on quantum computing's software and system level, providing an essential learning platform for novices and seasoned practitioners alike. The introductory session aims to equip individuals with the foundational knowledge required to navigate the complex field of quantum computing. This initial phase paves the way for participants to appreciate the significance of recent advancements and emerging research trends, which are highlighted in the subsequent research session. Notably, the software and system level-focused approach caters to the diverse backgrounds of our audience. It creates an avenue for those lacking relevant experience to learn quantum fundamentals while enabling practitioners to reinforce their existing knowledge. Additionally, the research topic session fosters a dynamic environment for discussions on the latest research trends, encouraging collaborative interactions with experts.

The paper is structured as follows: Section II outlines the lecture design and introduces confirmed speakers, while Section III shares insights from the first completed lecture session. Finally, Section IV concludes with a comprehensive summary and future expectations. The presented lecture series represents a bold step towards enhancing understanding and fostering global engagement in the dynamic field of quantum computing.

\section{Motivation and Expectation}
\subsection{Motivation}
Traditional quantum computing lecture series and courses have a tendency to emphasize the design of quantum hardware devices, related theoretical physics, and the mathematical aspects of quantum information processing. This leaves the quantum software and system level relatively unexplored. In our review of the existing literature and course offerings, we found that although some institutions offer courses on quantum software and systems, a significant number do not. Online resources on the subject are also scarce. Furthermore, given that quantum software and system level is an emerging field, there are only a handful of professors specializing in it, resulting in a dearth of learning resources for interested students.

\subsection{Expectation}
Our aim is to curate a lecture series that initiates with the basics of quantum computing software and systems, empowering audiences to comprehend, use, and design software and systems for quantum computing. Additionally, we intend to incorporate discussions of cutting-edge research to provide the audience with a deeper understanding and a propensity towards research topics. We have established several expectations for the lecture series:

\begin{itemize}
    \item \textbf{Foundation for Beginners:} The first part of the lecture is an introductory session, aimed at audience members who have an interest in quantum computing software and system level but lack a background in it. Topics covered include algorithms, quantum compilation, noise mitigation, quantum machine learning, quantum architecture design, quantum simulation, quantum networking, among others.
    \item \textbf{Deepening Understanding for Researchers:} The software and systems aspects of quantum computing represent a rapidly evolving field. We anticipate that even experienced researchers will gain fresh insights into the basics from the varied perspectives of different experts in the field.
    
    \item \textbf{Exploring Research Frontiers:} After laying the foundational knowledge, the lecture series transitions to the research topic session. More than 35 senior researchers in the field have confirmed their participation, each presenting their respective research topics. We believe that engaging in discussions on these cutting-edge topics is instrumental for the progress of the field. We expect that not only will the audience draw inspiration from the presentations, but the speakers will also identify potential research gaps or receive valuable feedback during the Q \& A session. 
    \item \textbf{Increase the Diversity of the Field.} In an attempt to create a diverse learning environment and address the underrepresentation within the quantum computing sector, the lecture series is designed to engage individuals from varying academic backgrounds and regions. Encouraging global participation and cross-cultural collaboration, we hope to bring unique perspectives, fostering creativity and driving innovation in the quantum software and system level sphere. We anticipate that the lecture series will act as a catalyst for nurturing a global community of quantum computing researchers and practitioners, expanding the reach of quantum computing knowledge, and fostering a diverse and inclusive research ecosystem. 
    Moreover, we aim to enhance the presence of underrepresented groups within the quantum software and system level field, believing that diversity in contributors' backgrounds can bring about robust and multi-faceted approaches to problem-solving. The series will incorporate the efforts to break down barriers and actively engage more women, people of color, and individuals from other marginalized groups in the quantum computing community. Ultimately, we strive to build a more diverse and inclusive field, which we believe will be a key driver in accelerating the pace of innovation in quantum computing.
\end{itemize}

\section{Quantum Computer System Lecture Series}
\label{sec2}
This lecture series is named Quantum Computer System Lecture Series (QuCS). It is organized by Zhiding Liang, a PhD student at the University of Notre Dame, and co-organized by Hanrui Wang, a PhD student at MIT. Confirmed over 40 speakers include professors and senior Ph.D. students from 27 universities (e.g., MIT, Princeton, Yale, Harvard, USC, TUM, Oxford, UChicago, Montpelier) and researchers from industry (e.g., IBM, JPMorgan Chase \& Co, and Wells Fargo). They are geographically located in eight countries: the United States, China, England, France, Australia, Vietnam, Germany, and Canada. Our lecture series starts in July 2022 and is divided into introduction sessions and research topic sessions. The first session is held on July 7, 2022 between 10:30 Am and 11:30 Am EST on Zoom. The time is selected to accommodate the audience in Europe, Asia as well as North/South America. Most of the subsequent lectures will be held at the same time period (10:30 Am - 11:30 Am EST) on Thursdays, with each lecture divided into 45-50 minutes for the speaker to talk about the main topic and 10-15 minutes for the follow-up Question \& Answer.

\subsection{Schedule}
The updated schedule of the QuCS lecture series showing as follows:

\textbf{Session 1: Introduction Session}
\vspace{2mm}

\noindent
\fbox{%
\begin{minipage}{0.45\textwidth}
\begin{description}
    \item[Name:] Prof. Yongshan Ding
    \item[Title:] Assistant Professor at Yale University
    \item[Topic:] Software and Algorithmic Approaches to Quantum Noise Mitigation: An Overview
\end{description}
\end{minipage}
\noindent
}
\fbox{%
\begin{minipage}{0.45\textwidth}
\begin{description}
    \item[Name:] Zhixin Song
    \item[Title:] PhD Student at Georgia Institute of Technology
    \item[Topic:] A Guided Tour on the Map of Quantum Computing
\end{description}
\end{minipage}
}
\noindent
\fbox{%
\begin{minipage}{0.45\textwidth}
\begin{description}
    \item[Name:] Jinglei Cheng
    \item[Title:] PhD Student at University of Southern California
    \item[Topic:] Introduction to Variational Quantum Algorithms
\end{description}
\end{minipage}
}
\noindent
\fbox{%
\begin{minipage}{0.45\textwidth}
\begin{description}
    \item[Name:] Siyuan Niu
    \item[Title:] PhD Candidate at University of Montpellier
    \item[Topic:] Enabling Parallel Circuit Execution on NISQ Hardware
\end{description}
\end{minipage}
}
\noindent
\fbox{%
\begin{minipage}{0.45\textwidth}
\begin{description}
    \item[Name:] Prof. Robert Wille
    \item[Title:] Distinguished Professor at Technische Universität München
    \item[Topic:] Design Automation and Software Tools for Quantum Computing
\end{description}
\end{minipage}
}
\noindent
\fbox{%
\begin{minipage}{0.45\textwidth}
\begin{description}
    \item[Name:] Prof. Tongyang Li
    \item[Title:] Assistant Professor at Peking University
    \item[Topic:] Adaptive Online Learning of Quantum States
\end{description}
\end{minipage}
}
\noindent
\fbox{%
\begin{minipage}{0.45\textwidth}
\begin{description}
    \item[Name:] Dr. Junyu Liu
    \item[Title:] Theoretical Physicist at the University of Chicago
    \item[Topic:] Quantum Data Center
\end{description}
\end{minipage}
}
\noindent
\fbox{%
\begin{minipage}{0.45\textwidth}
\begin{description}
    \item[Name:] Dr. Gokul Ravi
    \item[Title:] Computing Innovation Fellow at the University of Chicago
    \item[Topic:] Classical Support and Error Mitigation for Variational Quantum Algorithms
\end{description}
\end{minipage}
}
\noindent
\fbox{%
\begin{minipage}{0.45\textwidth}
\begin{description}
    \item[Name:] Prof. Chen Qian
    \item[Title:] Professor at UCSC
    \item[Topic:] Protocol Design for Quantum Network Routing
\end{description}
\end{minipage}
}
\noindent
\fbox{%
\begin{minipage}{0.45\textwidth}
\begin{description}
    \item[Name:] Yilian Liu
    \item[Title:] Ms student at Cornell University
    \item[Topic:] Solving Nonlinear Partial Differential Equations using Variational Quantum Algorithms on Noisy Quantum Computers
\end{description}
\end{minipage}
}

\vspace{5mm}
\textbf{Session 2: Research Session}

\noindent
\fbox{%
\begin{minipage}{0.45\textwidth}
\begin{description}
    \item[Name:] Prof. Prineha Narang
    \item[Title:] Associate Professor at UCLA
    \item[Topic:] Building Blocks of Scalable Quantum Information Science
\end{description}
\end{minipage}
}
\noindent
\fbox{%
\begin{minipage}{0.45\textwidth}
\begin{description}
    \item[Name:] Prof. Jakub Szefer
    \item[Title:] Associate Professor at Yale University
    \item[Topic:] Quantum Computer Hardware Cybersecurity
\end{description}
\end{minipage}
}
\noindent
\fbox{%
\begin{minipage}{0.45\textwidth}
\begin{description}
    \item[Name:] Prof. Guan Qiang
    \item[Title:] Assistant Professor at Kent State University
    \item[Topic:] Enabling robust quantum computer system by understanding errors from NISQ machines
\end{description}
\end{minipage}
}
\noindent
\fbox{%
\begin{minipage}{0.45\textwidth}
\begin{description}
    \item[Name:] Bochen Tan
    \item[Title:] PhD student at UCLA
    \item[Topic:] Compilation for Near-Term Quantum Computing: Gap Analysis and Optimal Solution
\end{description}
\end{minipage}
}
\noindent
\fbox{%
\begin{minipage}{0.45\textwidth}
\begin{description}
    \item[Name:] Zeyuan Zhou
    \item[Title:] PhD student at JHU
    \item[Topic:] Quantum Crosstalk Robust Quantum Control
\end{description}
\end{minipage}
}
\noindent
\fbox{%
\begin{minipage}{0.45\textwidth}
\begin{description}
    \item[Name:] Wei Tang
    \item[Title:] PhD student at Princeton
    \item[Topic:] Distributed Quantum Computing
\end{description}
\end{minipage}
}
\noindent
\fbox{%
\begin{minipage}{0.45\textwidth}
\begin{description}
    \item[Name:] Prof. Tirthak Patel
    \item[Title:] Assistant Professor at Rice University
    \item[Topic:] Developing Robust System Software Support for Quantum Computers
\end{description}
\end{minipage}
}
\noindent
\fbox{%
\begin{minipage}{0.45\textwidth}
\begin{description}
    \item[Name:] Prof. Gushu Li
    \item[Title:] Assistant Professor at the University of Pennsylvania
    \item[Topic:] Enabling Deeper Quantum Compiler Optimization at High Level
\end{description}
\end{minipage}
}
\noindent
\fbox{%
\begin{minipage}{0.45\textwidth}
\begin{description}
    \item[Name:] Prof. Nai-Hui Chia
    \item[Title:] Assistant Professor at the Rice University
    \item[Topic:] Classical Verification of Quantum Depth
\end{description}
\end{minipage}
}
\noindent
\fbox{%
\begin{minipage}{0.45\textwidth}
\begin{description}
    \item[Name:] Prof. Mohsen Heidari
    \item[Title:] Assistant Professor at the Indiana University, Bloomington
    \item[Topic:] Learning and Training in Quantum Environments
\end{description}
\end{minipage}
}
\noindent
\fbox{%
\begin{minipage}{0.45\textwidth}
\begin{description}
    \item[Name:] Prof. Jun Qi
    \item[Title:] Assistant Professor at the Fudan University
    \item[Topic:] Quantum Machine Learning: Theoretical Foundations and Applications on NISQ Devices
\end{description}
\end{minipage}
}
\noindent
\fbox{%
\begin{minipage}{0.45\textwidth}
\begin{description}
    \item[Name:] Yasuo Oda
    \item[Title:] PhD student at John Hopkins University
    \item[Topic:] Noise Modeling of the IBM Quantum Experience
\end{description}
\end{minipage}
}
\noindent
\fbox{%
\begin{minipage}{0.45\textwidth}
\begin{description}
    \item[Name:] Yuxiang Peng
    \item[Title:] PhD student at the University of Maryland, College Park
    \item[Topic:] Software Tools for Analog Quantum Computing
\end{description}
\end{minipage}
}
\noindent
\fbox{%
\begin{minipage}{0.45\textwidth}
\begin{description}
    \item[Name:] Runzhou Tao
    \item[Title:] PhD student at Columbia University
    \item[Topic:] Automatic Formal Verification of the Qiskit Compiler
\end{description}
\end{minipage}
}
\noindent
\fbox{%
\begin{minipage}{0.45\textwidth}
\begin{description}
    \item[Name:] Jiyuan Wang
    \item[Title:] PhD Candidate at University of California, Los Angeles
    \item[Topic:] QDiff: Differential Testing for Quantum Software Stacks
\end{description}
\end{minipage}
}
\noindent
\fbox{%
\begin{minipage}{0.45\textwidth}
\begin{description}
    \item[Name:] Charles Yuan
    \item[Title:] PhD student at Massachusetts Institute of Technology
    \item[Topic:] Abstractions Are Bridges Toward Quantum Programming
\end{description}
\end{minipage}
}
\noindent
\fbox{%
\begin{minipage}{0.45\textwidth}
\begin{description}
    \item[Name:] Minzhao Liu
    \item[Title:] PhD student at University of Chicago
    \item[Topic:] Understanding Quantum Supremacy Conditions for Gaussian Boson Sampling with High Performance Computing
\end{description}
\end{minipage}
}
\noindent
\fbox{%
\begin{minipage}{0.45\textwidth}
\begin{description}
    \item[Name:] Dr. Naoki Kanazawa
    \item[Title:] Research Scientist at IBM Quantum
    \item[Topic:] Pulse Control for Superconducting Quantum Computers
\end{description}
\end{minipage}
}
\noindent
\fbox{%
\begin{minipage}{0.45\textwidth}
\begin{description}
    \item[Name:] Prof. He Li
    \item[Title:] Associate Professor at Southeast University (China)
    \item[Topic:] Rethinking Most-significant Digit-first Arithmetic for Quantum Computing in NISQ Era
\end{description}
\end{minipage}
}
\noindent
\fbox{%
\begin{minipage}{0.45\textwidth}
\begin{description}
    \item[Name:] Dr. Thinh DINH
    \item[Title:] Researcher at Vietnam National University
    \item[Topic:] Efficient Hamiltonian Reduction for Scalable Quantum Computing on Clique Cover/Graph Coloring Problems in SatCom
\end{description}
\end{minipage}
}
\noindent
\fbox{%
\begin{minipage}{0.45\textwidth}
\begin{description}
    \item[Name:] Dr. Marco Pistoia
    \item[Title:] Managing Director, Distinguished Engineer, and Head of JPMorgan Chase’s Global Technology Applied Research Center
    \item[Topic:] Quantum Computing and Quantum Communication in the Financial World
\end{description}
\end{minipage}
}
\noindent
\fbox{%
\begin{minipage}{0.45\textwidth}
\begin{description}
    \item[Name:] Prof. Yuan Feng
    \item[Title:] Professor at the Centre for Quantum Software and Information, University of Technology Sydney
    \item[Topic:] Hoare logic for verification of quantum programs
\end{description}
\end{minipage}
}
\noindent
\fbox{%
\begin{minipage}{0.45\textwidth}
\begin{description}
    \item[Name:] Lia Yeh
    \item[Title:] PhD student at University of Oxford
    \item[Topic:] Quantum Graphical Calculi: Tutorial and Applications
\end{description}
\end{minipage}
}
\noindent
\fbox{%
\begin{minipage}{0.45\textwidth}
\begin{description}
    \item[Name:] Zhirui Hu
    \item[Title:] PhD student at George Mason University
    \item[Topic:] Optimize Quantum Learning on Near-Term Noisy Quantum Computers
\end{description}
\end{minipage}
}
\noindent
\fbox{%
\begin{minipage}{0.45\textwidth}
\begin{description}
    \item[Name:] Prof. Zhu Han
    \item[Title:] John and Rebecca Moores Professor at University of Houston
    \item[Topic:] Hybrid Quantum-Classic Computing for Future Network Optimization
\end{description}
\end{minipage}
}
\noindent
\fbox{%
\begin{minipage}{0.45\textwidth}
\begin{description}
    \item[Name:] Dr. Samuel Yen-Chi Chen
    \item[Title:] Senior Software Engineer at Wells Fargo
    \item[Topic:] Hybrid Quantum-Classical Machine Learning with Applications
\end{description}
\end{minipage}
}
\noindent
\fbox{%
\begin{minipage}{0.45\textwidth}
\begin{description}
    \item[Name:] Dr. Ji Liu
    \item[Title:] Postdoctoral Researcher at Argonne National Laboratory
    \item[Topic:] Elevating Quantum Compiler Performance through Enhanced Awareness in the Compilation Stages
\end{description}
\end{minipage}
}
\noindent
\fbox{%
\begin{minipage}{0.45\textwidth}
\begin{description}
    \item[Name:] Dr. Daniel Egger
    \item[Title:] Senior Research Scientist at IBM Quantum, IBM Zurich
    \item[Topic:] Pulse-based Variational Quantum Eigensolver and Pulse-Efficient Transpilation
\end{description}
\end{minipage}
}
\noindent
\fbox{%
\begin{minipage}{0.45\textwidth}
\begin{description}
    \item[Name:] Thomas Alexander
    \item[Title:] Software Developer at IBM Quantum, Market Leader in Quantum Systems and Services
    \item[Topic:] Control Systems \& Systems Software at IBM Quantum
\end{description}
\end{minipage}
}
\begin{figure*}[t]
\centering
\includegraphics[width=\linewidth]{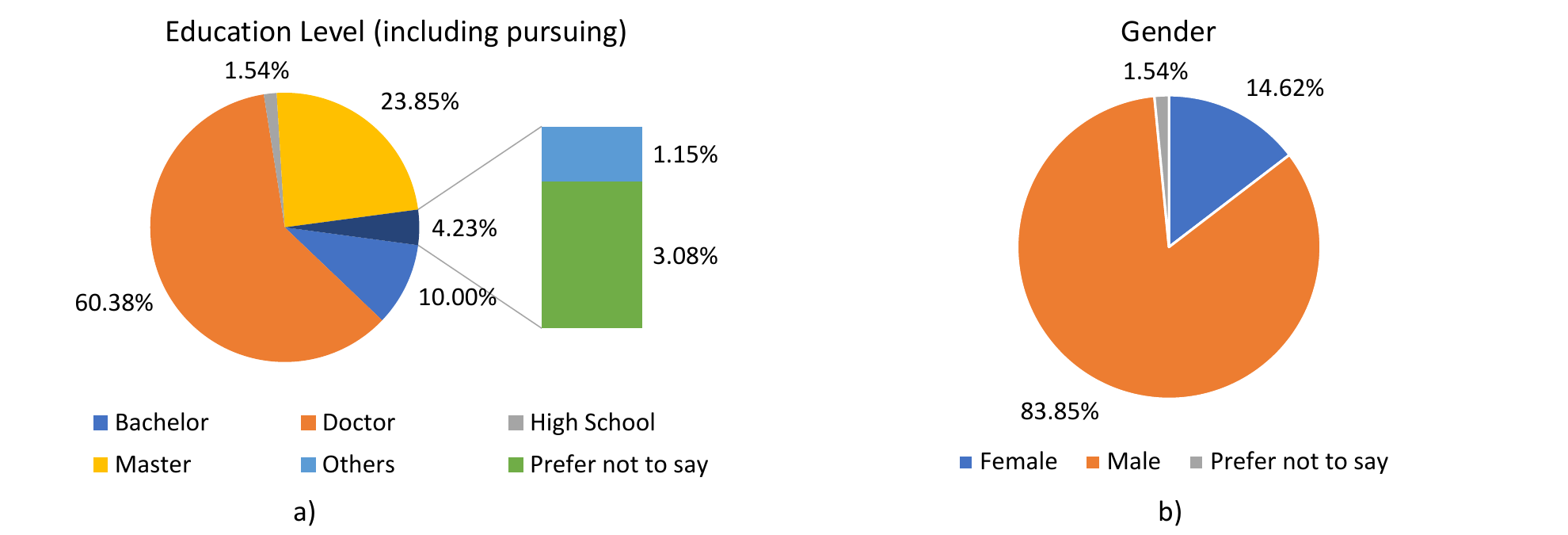}
\caption{Background survey data for a) Education level, and b) Gender of 260 participants.}
\label{education}
\end{figure*}

\subsection{General Data}

Our lectures currently have 1294 registered audience, and this number is still increasing. 
So far we have invited speakers from 27 different universities in seven different countries to cover the fundamentals of quantum computing at software and system level as well as the frontier topics.

\subsection{Collaboration with Industry Experts}

In our pursuit of comprehensive knowledge sharing and insightful discussions, we have reached beyond the realm of academia. We are delighted to announce the participation of distinguished experts from renowned corporations such as IBM, JPMorgan Chase \& Co., and Wells Fargo. This fusion of academic and industry perspectives is intended to equip our audience with a holistic understanding of quantum computing, particularly in the context of quantum software and system level.

These industry professionals bring valuable insights from the forefront of quantum computing applications, highlighting not only theoretical aspects but also practical considerations. Our belief is that the integration of academia and industry will foster a dynamic and enriching learning environment. By understanding both theoretical underpinnings and industrial applications, our audience will gain a well-rounded grasp of quantum computing at the software and system level. This not only broadens the scope of knowledge but also elucidates the potential and challenges of quantum computing in the contemporary business landscape. We hope that this rich blend of insights will spark innovative thinking and inspire the next wave of advancements in the quantum computing arena.

\section{Observation}
Based on the collected data from our background and detailed surveys, we have gleaned a range of insights regarding our lecture participants. The background survey had 260 participants, while the detailed survey was filled out by 28 individuals.

\subsection{Background Survey Observations}

The background survey encompassed three main areas: education level, gender, and region. Out of the 260 participants, a significant 60.38\% have achieved or are in the process of obtaining their doctoral degree as shown in the Fig. \ref{education}(a). This large representation reflects the continued and strong interest in quantum computing from research-focused individuals. Those who have achieved or are pursuing a Master's degree and a Bachelor's degree each constituted 23.85\% and 10.00\% of the participants, respectively. These numbers confirm that the appeal of quantum computing is permeating all levels of academia.

Surprisingly, 1.54\% of participants were high school graduates or high school students. This indicates that quantum computing has started to attract the attention of the K-12 education group to a certain extent. Such a trend is an encouraging sign for the future of the quantum computing community, suggesting an expansion of interest beyond the traditional confines of higher education and research.

When we analyzed the gender representation among participants, as shown in the Fig. \ref{education}(b), we found that a significant portion, 83.85\%, identified as male, either biologically or psychologically. Conversely, the percentage of those who identify as female is considerably smaller, at just 14.62\%. These figures present a substantial gender imbalance within the community of our lecture attendees. They suggest that much work needs to be done to make quantum computing more appealing and accessible to a broader, more diverse audience. Specifically, it highlights the urgency for initiatives within the quantum computing community to foster a more inclusive environment that actively encourages participation from minorities.

In this light, the QuCS series is committed to taking steps towards this goal, by fostering a more welcoming and diverse community, amplifying the voices of minority speakers, and presenting topics relevant to a wider audience, among other strategies.

\subsection{Detailed Survey Observations}

\begin{table*}[]
  \captionsetup{}
  \setlength{\tabcolsep}{2.5em}
\begin{tabular}{|c|c|c|}
\hline
Knowledge Level & Description                                                                                  & Percentage \\ \hline
Level 0         & I am interested but don't know anything.                                                     & 17.9\%     \\ \hline
Level 1         & I used to study by myself for months or take one course related but still not familiar.      & 32.1\%     \\ \hline
Level 2         & I understand the quantum basis and have ability to finish a course-level project by my self. & 14.3\%     \\ \hline
Level 3         & I am familiar with at least one area in quantum computing.                                   & 32.1\%     \\ \hline
Level 4         & I am an expert of quantum computing.                                                         & 3.6\%      \\ \hline
\end{tabular}
\caption{Description of the defined knowledge level of participants about quantum computing.}
\label{levels}
\end{table*}

\begin{figure*}[t]
\centering
\includegraphics[width=\linewidth]{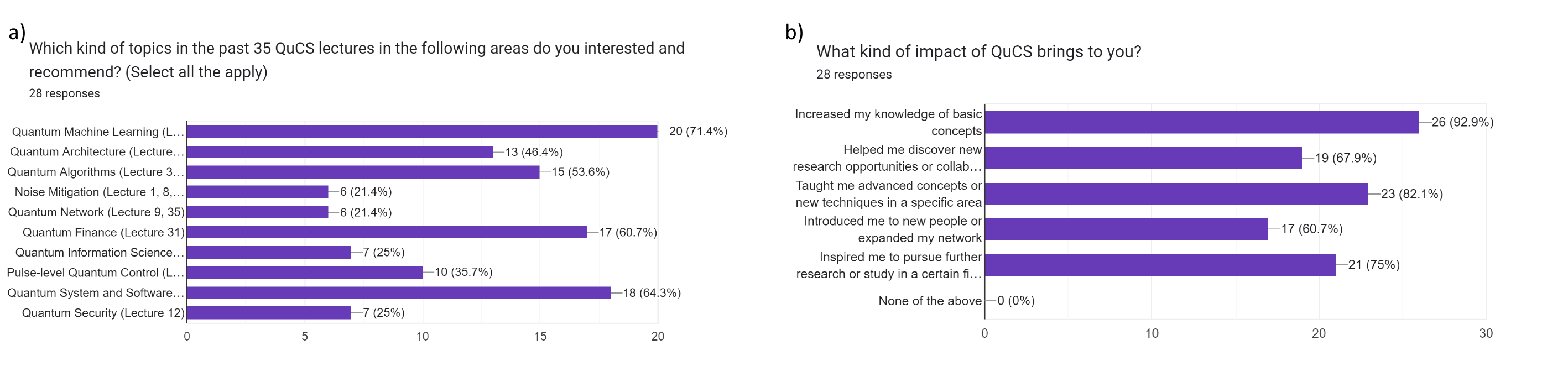}
\caption{a) Voting counts for the most interesting topic among the past 35 lectures at QuCS. b) Voting counts for the impact of the QuCS brings to audiences.}
\label{topic}
\end{figure*}

\begin{figure}[t]
\centering
\includegraphics[width=\linewidth]{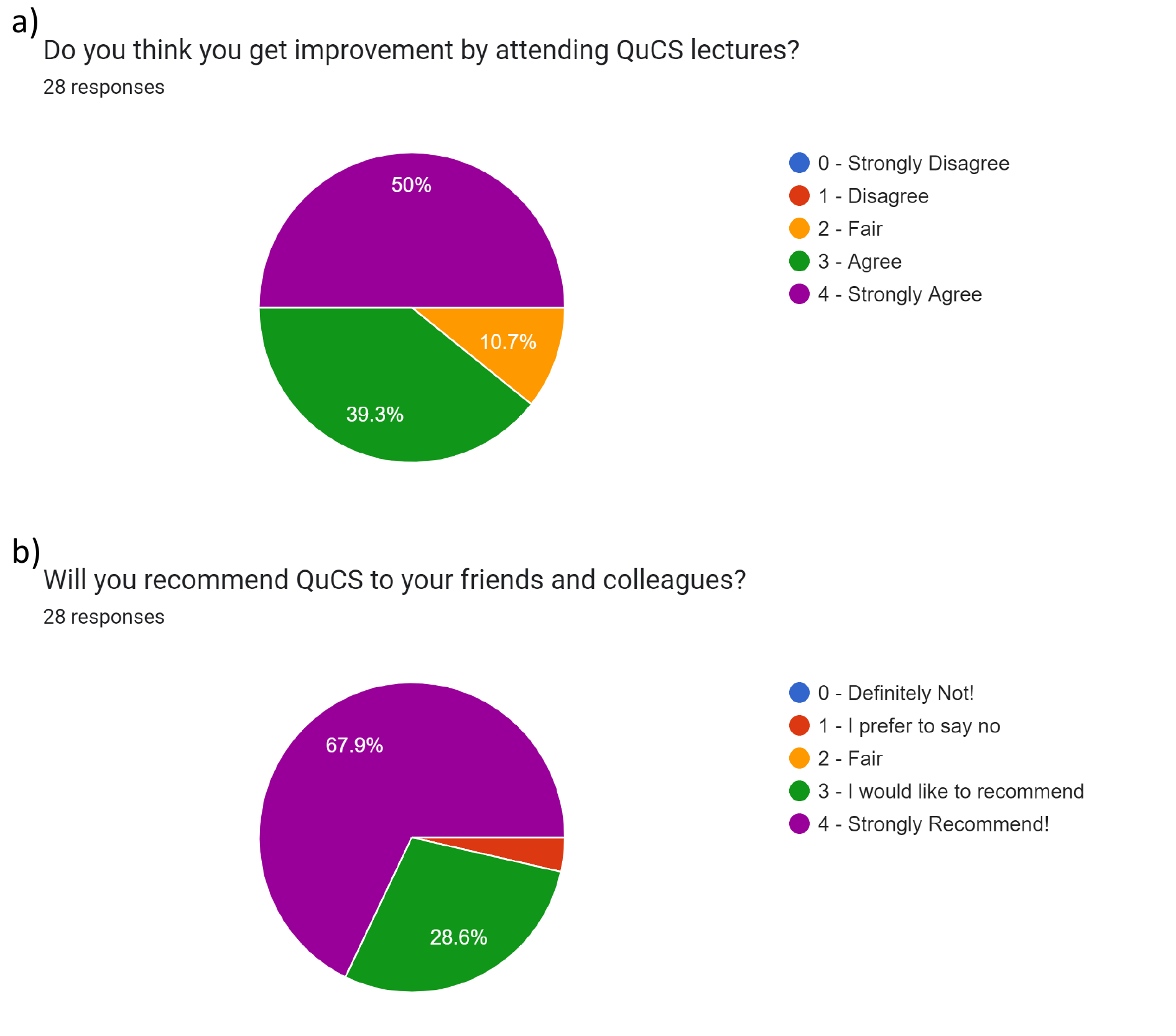}
\caption{Feedback data: a) if audiences gain improvement from QuCS. b) will audiences recommend QuCS to friends and colleagues.}
\label{eva}
\end{figure}

\begin{figure}[t]
\centering
\includegraphics[width=\linewidth]{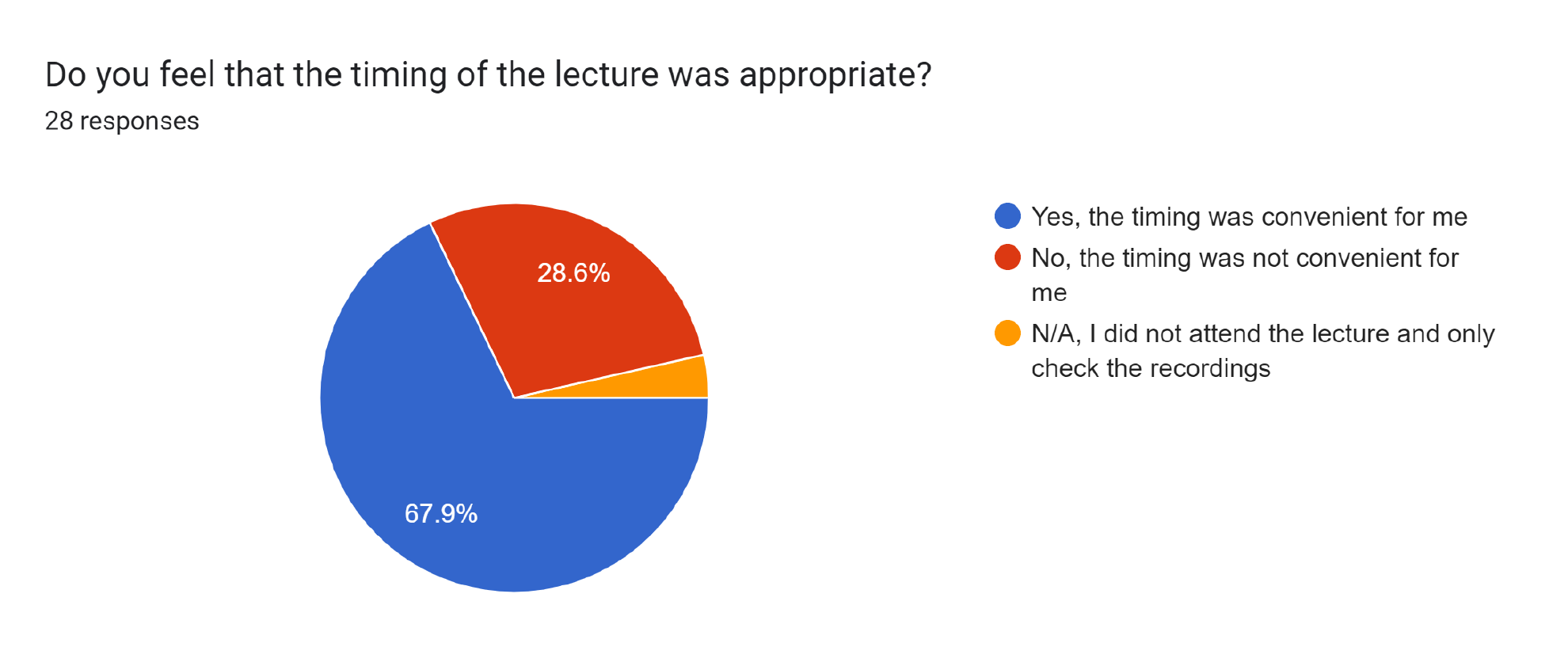}
\caption{Feedback data: if the timing of QuCS is appropriate for audiences.}
\label{time}
\end{figure}

The detailed survey covered several areas, including participants' perceived knowledge level of quantum computing, their principal job function, ethnicity, race, underrepresented group identification, topics of interest, knowledge about QuCS, the perceived impact of QuCS lectures, lecture timing appropriateness, and whether they would recommend QuCS. We analyze this data in the subsequent sections to further understand our audience's learning experience, preferences, and overall impact of the QuCS lecture series. We also delve into a more detailed analysis of the information gathered from these surveys to further comprehend the effect of our lecture series and how we can adapt to cater to our audience's needs better.

\textbf{Knowledge Level: }In terms of self-reported knowledge level in quantum computing among detailed survey respondents, we observed that a significant portion of the respondents, precisely 50\%, considered themselves to be at a beginner level (Levels 0 - 1). These participants suggested that their understanding of quantum computing concepts was still in the early stages. Meanwhile, 14.3\% of the respondents believed that they had achieved Level 2 proficiency, indicating that they could independently complete a course project related to quantum computing. This level is an encouraging sign of the participants' commitment to developing their skills further. Simultaneously, a notable percentage (32.1\%) of respondents indicated that they have substantial expertise in a specific area of quantum computing. Lastly, a small fraction of the respondents, at 3.6\%, saw themselves as highly knowledgeable experts in the quantum computing field. These individuals likely have substantial experience and engagement in the field, possibly contributing to research and development in the area. This distribution of self-assessed proficiency levels indicates a diverse audience in terms of knowledge and experience. 

\textbf{Attractive Topics: }The detailed survey also provided insights into the participants' interests and recommendations regarding the topics covered in the previous 35 QuCS lectures as shown in Fig. \ref{topic}(a). Quantum machine learning emerged as the most intriguing topic, with 20 out of 28 respondents expressing significant interest. Quantum system and software design and quantum finance were also notable areas of interest, with 18 and 17 votes respectively. Quantum algorithms, quantum architecture, and quantum control received moderate interest, each gathering between 10 to 15 votes. Topics with lower engagement levels were quantum networking and quantum security, suggesting potential areas for enhanced focus and engagement in future lectures. 

\textbf{Impact of QuCS: }The impact of the QuCS lecture series was another area of interest in the detailed survey. As shown in the Fig. \ref{topic}(b), a strong majority of respondents (26 out of 28) indicated that the lectures had significantly enhanced their understanding of basic quantum computing concepts, aligning well with one of the key objectives of QuCS, i.e., to strengthen the audience's background in quantum computing. Moreover, 23 respondents believed that QuCS had exposed them to advanced research topics, while 21 felt that their scientific interest in certain research topics was stimulated due to their involvement with QuCS. This aligns well with the intentions behind the establishment of the research topics session within QuCS.

Furthermore, the QuCS lectures seemed to have made a contribution to building a more robust quantum computing community. Specifically, 19 participants stated that the series had helped them find new collaboration partners, and 17 believed it had assisted in creating new social networks for them. This reflects positively on the central efforts of QuCS to foster connections and collaboration within the quantum computing community. These findings underscore the value of QuCS and its potential to continue making significant contributions to the field in the future.

\textbf{Feedback from Audience: }Moving on to the overall feedback about the QuCS series, we have analyzed responses to three key questions: 
\begin{itemize}
    \item 1) Do you think you get improvement by attending QuCS lectures?
    \item 2) Will you recommend QuCS to your friends and colleagues?
    \item 3) Do you feel that the timing of the lecture was appropriate?
\end{itemize}

The data collected from the first question was highly encouraging, with 89.3\% of participants choosing either 'agree' or 'strongly agree'. This implies that a majority of the participants felt they had greatly benefited from the QuCS lectures. However, 10.7\% of respondents chose 'fair', possibly indicating that the breadth of topics covered in QuCS is not yet sufficient. Some suggestions were made to include lectures on quantum error correction and quantum optimization.

For the second question, 27 out of 28 respondents indicated they would 'recommend' or 'strongly recommend' QuCS to their friends and colleagues, while one participant did not wish to share this information.

Lastly, 67.9\% of participants felt that the timing of the QuCS lectures was suitable. Timing is a significant challenge when organizing these lectures, as we aim to accommodate audiences and speakers from different time zones. In setting the lecture time, we mainly considered the time zones of Asia, Europe, and North America, scheduling them for 10:30 a.m. Eastern Time. In the future, we plan to conduct a detailed survey regarding the most suitable lecture time and reevaluate our current schedule accordingly.

\section{Conclusion and Future Work}
\label{sec5}
As the field of quantum computing is gaining importance and the need for a quantum workforce in the industry is becoming more urgent, and the need for quantum computing faculty in CS and CE majors in academia is gradually increasing, the expansion beyond the roots in physical and theoretical becoming the next goal of quantum computing education. Our organized QuCS aims to provide open educational opportunities in quantum computing on software and systems to participants with diverse quantum computing knowledge backgrounds. We offer introductory participants the opportunity to learn fundamental knowledge of quantum computing on software and systems, and participants already working in the field the opportunity to refresh their knowledge and gain new understanding. We also provide opportunities for all audience members to discuss research topics. And we have designed QuCS from introductory to in-depth, and have made all lecture recordings publicly available to ensure that all are given the opportunity to build a relevant background of knowledge before discussing the research topic.

We are devoted to expanding the diversity of the quantum computing research community.  And we invite as well as organize speaker and participants from different universities, different countries, different races, and different genders.

We will continue make contribution to the education on software and system on quantum computing, and keeping to invite experts to join our lecture series. Also, we plan to expand our lecture series to a workshop in quantum computing/ architecture/ design automation/ machine learning conference. We expect that on day we will have a chance to invite speakers that who began in this field inspired by our lecture series.

\section*{Acknowledgment}
We thank all speakers and audiences who were interested and participated in our lecture series.

\printbibliography
\end{document}